\documentstyle[12pt]{article}
\input{epsf}
\newcommand{\be}{\begin{eqnarray}}
\newcommand{\ee}{\end{eqnarray}}
\newcommand{\ba}{\begin{array}}
\newcommand{\ea}{\end{array}}
\newcommand{\bftau}{\mbox{\boldmath{$\tau$}}}
\begin{document}
\setcounter{page}{0}
\thispagestyle{empty}
\rightline{RUB-TPII-12/97}
\rightline{hep-ph/9712447}
\vspace{.3cm}
\begin{center}
\begin{large}
{\bf Self-consistent calculation of parton distributions at low 
normalization point in the chiral quark--soliton model} 
\\[2cm]
\end{large}
{\bf C. Weiss$^{\rm a}$ and K. Goeke$^{\rm b}$}
\\
{\em Institut f\"ur Theoretische Physik II,
Ruhr--Universit\"at Bochum, \\ D--44780 Bochum, Germany} 
\end{center}
\vspace{1.5cm}
\begin{abstract}
\noindent
We calculate the isosinglet unpolarized quark-- and antiquark distributions
at a low normalization point in the large--$N_c$ limit. The nucleon is
described as a self-consistent soliton solution of the effective chiral
theory.  The ultraviolet cutoff is implemented by a Pauli--Villars
subtraction. The quark and antiquark distributions satisfy the momentum as
well as the baryon number sum rule, corresponding to ``constituent'' quarks
whose structure is not yet resolved at the low scale.  Already at this
level a sizable number of antiquarks is present.
\end{abstract}
\vfill
\rule{5cm}{.15mm}
\\
\noindent
{\footnotesize $^{\rm a}$ E-mail: weiss@hadron.tp2.ruhr-uni-bochum.de} \\
{\footnotesize $^{\rm b}$ E-mail: goeke@hadron.tp2.ruhr-uni-bochum.de}
\newpage
\noindent
%
%
An essential ingredient in the theory of deep--inelastic scattering and
other hard processes are the leading--twist parton distribution functions
of the nucleon \cite{MRS95,GRV95}. Their evolution in the asymptotic region
is governed by perturbative QCD and well understood. However, the initial
conditions for the perturbative evolution, {\em i.e.}, the parton
distributions at a relatively low normalization point, belong to the field
of non-perturbative physics, and at present one has to rely on model
calculations to estimate them.
\par
Recently, an approach has been formulated to calculate the leading--twist
parton distributions in the large--$N_c$ limit, where the nucleon can be
described as a chiral soliton \cite{DPPPW96,DPPPW97}. It was shown that
this field--theoretic approach allows to preserve all general requirements
on parton distributions such as positivity and the partonic sum rules which
hold in QCD. Results for the leading quark-- and antiquark distribution
function in the $1/N_c$--expansion (isosinglet unpolarized and isovector
polarized) show reasonable agreement with the parametrizations of the data
at a low normalization point \cite{GRV95}; the approach has been extended
to transverse polarized distributions \cite{PP96}. First calculations of
the subleading distributions have also been reported \cite{Wakamatsu97}.
\par
The description of the nucleon as a chiral soliton is based on the
dynamical breaking of chiral symmetry, which can be encoded in an effective
theory, valid at low energies. The effective theory can be derived from the
instanton vacuum of QCD, in which case the ultraviolet cutoff is determined
by the inverse average instanton size, $\bar\rho^{-1} = 600\,{\rm MeV}$,
the scale at which the dynamically generated ``constituent'' quark mass
drops to zero \cite{DP86}.  In actual calculations the fall--off of the
dynamical quark mass is usually simulated by applying some external UV
regularization. Hadronic observables, such as the nucleon mass, axial
coupling constant {\em etc.}\ are, as a rule, insensitive to the details of
the UV cutoff. When one turns to the calculation of parton distribution
functions, however, there are very strong restrictions on how one should
introduce the UV cutoff in the effective theory. The point is that one has
to preserve certain general properties of parton distributions such as
positivity, sum rules {\em etc.}, which can easily be violated by an
arbitrary UV regularization.  Specifically, the regularization should
preserve the completeness of the set of quark single--particle wave
functions in the soliton pion field. One possible regularization method
which fulfills all requirements is a Pauli--Villars subtraction
\cite{DPPPW96,DPPPW97}.
\par
An important aspect which has not been taken into account in the
calculation of parton distribution functions so far, related to the
constraints on the UV regularization, is the self--consistent description
of the chiral soliton. The classical pion field describing the nucleon
should be determined as a minimum of the static energy, suitably
regularized in a way that ensures consistency with the regularization
applied to the parton distributions. The point where the two
regularizations meet is the momentum sum rule for the isosinglet
distribution, which is satisfied only if the classical pion field is a
solution of the equations of motion of the effective chiral theory
\cite{DPPPW96}. The calculations of parton distributions in
refs.\cite{DPPPW96,DPPPW97} were performed with a particular variant of the
Pauli--Villars regularization and using a fixed soliton profile, and the
momentum sum rule was satisfied only approximately.
\par
In the present letter we compute the isosinglet unpolarized
distributions in the approach formulated in refs.\cite{DPPPW96,DPPPW97} with
a self--consistent description of the nucleon. In fact, a class of
regularizations exist which preserve all general requirements on parton
distributions (positivity, sum rules {\em etc}.) and simultaneously lead to
a stable minimum of the energy. With such a regularization the
momentum sum rule is explicitly preserved (along with other sum rules).
We also investigate the effects of possible finite regularizations
on the valence and, particularly, the antiquark distribution.
\par
The calculation of parton distributions from the effective chiral theory
relies on the parametric smallness of the ratio of the dynamical quark
mass, $M$, to the UV cutoff, $\Lambda$. [In the instanton vacuum this
ratio, $(M\bar\rho )^2$, is proportional to the small packing fraction of
instantons \cite{DP86}.]  The basic expressions for the twist--2 parton
distributions in refs.\cite{DPPPW96,DPPPW97} were derived in the limit
$M/\Lambda \rightarrow 0$. Differences between various UV regularizations
are parametrically of order $(M/\Lambda )^2$.  Since the existence of a
minimum of the energy depends on the details of the UV regularization, {\em
i.e.}, on terms of order $(M/\Lambda )^2$, one does not, from a principal
point of view, achieve a higher accuracy of the parton distributions by
making the description of the soliton self--consistent. Nevertheless, since
the existence of a stable soliton is an important qualitative issue --- not
only for the momentum sum rule --- it makes sense to ``improve'' the
regularization at level $(M/\Lambda )^2$ in order to ensure soliton
stability, and this is the point of view we take here.
\par
{\em Self-consistent soliton with Pauli--Villars regularization.}  The
classical pion field characterizing the nucleon in the effective chiral
theory at large $N_c$ is of ``hedgehog'' form \cite{DPP88},
\be
U ({\bf x}) &=& \exp\left[ i ({\bf n} \cdot \bftau ) P(r) \right]
\;\; = \;\; \cos P(r) + i ({\bf n} \cdot \bftau ) \sin P(r)
\nonumber \\
r &=& |{\bf x}|, \hspace{1cm} {\bf n} \; = \; \frac{{\bf x}}{r} .
\label{hedge}
\ee
The profile function, $P(r)$, is determined by minimizing the static energy
of the pion field,
\be
E_{\rm tot} [U] &=&  N_c E_{\rm lev} \; + \; N_c 
\sum_{\rm neg.\, cont.} ( E_n - E_n^{(0)} ) ,
\label{E_tot}
\ee
where $E_n$ are the quark single--particle energies, given as the
eigenvalues of the Dirac Hamiltonian in the background pion field ($M$ is
the dynamical quark mass),
\be
H\Phi_n &=& E_n \Phi_n ,
\hspace{1.5cm}
H \;\; = \;\; -i\gamma^0 \gamma^k \partial_k + M \gamma^0 U^{\gamma_5} ,
\nonumber \\
U^{\gamma_5}(x) &=& \frac{1+\gamma_5}2 U (x)
+ \frac{1-\gamma_5}2 U^\dagger (x).
\label{H} 
\ee
The spectrum of $H$ includes a discrete bound--state level, whose energy is
denoted by $E_{\rm lev}$, as well as the positive and negative Dirac
continuum, which are polarized by the presence of the pion field. The sum
in Eq.(\ref{E_tot}) runs over all occupied quark single--particle levels in
the soliton, that is, the bound--state level and the negative
continuum. The $E_n^{(0)}$ denote the energy levels of the vacuum
Hamiltonian given by Eq.(\ref{H}) with $U = 1$.
\par
The contribution of the Dirac continuum to the energy, Eq.(\ref{E_tot}), is
logarithmically divergent and requires UV regularization. We consider here
regularizations of the form of a Pauli--Villars subtraction, which can be
implemented consistently with the Pauli--Villars regularization of the
parton distribution functions below.  A possible regularization of
Eq.(\ref{E_tot}) is to subtract from the continuum contribution a suitable
multiple of the corresponding sum computed with the quark mass replace by a
regulator mass, $M_{PV}$:
\be
E_{\rm tot, reg} [U] &=&  N_c E_{\rm lev} 
\; + \; N_c \sum_{\rm neg.\, cont.} ( E_n - E_n^{(0)} ) 
\; - \; N_c \frac{M^2}{M_{PV}^2}
\sum_{\rm neg.\, cont.} ( E_{PV, n} - E_{PV, n}^{(0)} ) ,
\label{E_tot_reg}
\ee
where $E_{PV, n}$ are the eigenvalues of the Hamiltonian, Eq.(\ref{H}),
with $M$ replaced by $M_{PV}$ and the same pion field. The subtraction
coefficient follows from the fact that the logarithmic divergence of the
unregularized sum, Eq.(\ref{E_tot}), is proportional to $M^2$.  The value
of the regulator mass can be fixed from the (also logarithmically
divergent) pion decay constant,
\be
F_\pi^2 &=& 4N_c\int\frac{d^4k}{(2\pi)^4}\;\frac{M^2}{(M^2 + k^2)^2}
-4 N_c\frac{M^2}{M_{PV}^2}\int\frac{d^4k}{(2\pi )^4}\;
\frac{M_{PV}^2}{(M_{PV}^2 + k^2)^2} \nonumber \\
&=& \frac{N_c M^2}{4\pi^2} \log \frac{M_{PV}^2}{M^2} ,
\label{fpi}
\ee
$F_\pi = 93\, {\rm MeV}$. The regulator mass now plays the role of the UV
cutoff of the theory, $\Lambda$.
\par
It was shown in ref.\cite{Doering92} that the energy functional regularized
according to Eq.(\ref{E_tot_reg}) possesses a minimum with respect to the
profile function, {\em i.e.}, that a stable soliton exists. The minimum is
determined by the stationarity condition,
\be
\frac{\delta E_{\rm tot, reg}}{\delta P (r)} &=&
 -\sin P(r) {\cal S}(r) + \cos P(r) {\cal P}(r)
\;\; =\;\; 0 , 
\label{eom} 
\ee
where
\be
{\cal S}(r) &=& N_c {\cal S}_{\rm lev}(r) 
+ N_c \sum_{\rm neg.\, cont.} {\cal S}_{\rm n}(r)
- N_c \frac{M^2}{M_{PV}^2} \sum_{\rm neg.\, cont.} {\cal S}_{\rm PV, n}(r) ,
\\
{\cal P}(r) &=& N_c {\cal P}_{\rm lev}(r) 
+ N_c \sum_{\rm neg.\, cont.} {\cal P}_{\rm n}(r)
- N_c \frac{M^2}{M_{PV}^2} \sum_{\rm neg.\, cont.} {\cal P}_{\rm PV, n} (r) ,
\\
{\cal S}_n (r) &=& M \int d^3 x \; \Phi_n^\dagger ({\bf x}) \;
\delta (|{\bf x}| - r) \gamma^0 
\; \Phi_n ({\bf x}) , \\
{\cal P}_n (r) &=& M \int d^3 x \; \Phi_n^\dagger ({\bf x}) \; 
\delta (|{\bf x}| - r) \gamma^0 \gamma_5
i ({\bf n} \cdot \bftau ) \; \Phi_n ({\bf x}) ,
\ee
and similarly for ${\cal S}_{PV, n} (r), {\cal P}_{PV, n} (r)$.  The
profile function can be found by iterative solution of Eq.(\ref{eom}), see
ref.\cite{Doering92}, where also a number of hadronic nucleon observables
have been calculated with this regularization.  In the leading order of the
$1/N_c$--expansion the nucleon mass is given simply by the value of the
energy, Eq.(\ref{E_tot_reg}), at the minimum; with standard values for the
constituent quark mass, $M = 350\,{\rm MeV}\, (M_{PV}^2/M^2 = 2.25)$ and 
$M = 420\,{\rm MeV}\, (M_{PV}^2/M^2 = 1.90)$, we find, respectively, 
$M_N = 1140\,{\rm MeV}$ and $M_N = 1040\,{\rm MeV}$.
\par
The regularization of the energy, Eq.(\ref{E_tot_reg}), is different from
the one used in the calculations of refs.\cite{DPPPW96,DPPPW97}, where in
addition to the Dirac continuum also the contribution of the bound--state
level of the Hamiltonian with the regulator mass was subtracted. The latter
regularization does not lead to stable solitons. The two regularizations
differ by a finite contribution of order $M^2 / M_{PV}^2$. From the point
of view of computing parton distributions both regularizations belong to
the class of physically acceptable ones, see below.
\par
{\em Isosinglet unpolarized quark distribution function}. The theoretical
framework for the calculation of quark distribution functions with the
effective chiral theory has been given in refs.\cite{DPPPW96,DPPPW97}.  The
isosinglet unpolarized distribution appears in the leading order of the
$1/N_c$--expansion and can be represented as a sum of contributions of
quark single--particle levels in the background pion field:
\be
u(x) \, + \, d(x) &=& N_c f_{\rm lev}(x) \; + \; N_c 
\sum_{\rm neg.\, cont.} [ f_n (x) - f_n^{(0)}(x) ] , 
\label{q_uni} 
\\
f_n (x) &=& M_N \int\frac{d^3k}{(2\pi)^3}\;
\Phi_n^\dagger ({\bf k}) \; (1+\gamma^0\gamma^3) \, \delta(k^3 + E_n - xM_N)
\; \Phi_n ({\bf k}) .
\label{q_uni_level} 
\ee
This expression has been derived from the QCD representation of the parton
distribution as nucleon matrix element of a light--ray (non-local) operator
\cite{DPPPW96}; it is equivalent to the definition of the parton
distribution as the number of particles carrying a fraction $x$ of the
nucleon momentum in the infinite--momentum frame \cite{DPPPW97}.  The
function Eq.(\ref{q_uni}) describes the quark distribution for positive
values of $x$, and minus the antiquark distribution for negative $x$. In
Eq.(\ref{q_uni}) $f_n^{(0)}$ denotes the matrix element
Eq.(\ref{q_uni_level}) between levels of the vacuum Hamiltonian ($U = 1$),
with $f_n^{(0)}(x) = 0$ for $x > 0, E_n < 0$ on kinematical grounds.
\par
When computing the isosinglet distribution function in the effective theory
we must consider separately the even and odd part of the function defined
by Eq.(\ref{q_uni}), since they exhibit different behavior with respect to
the UV cutoff. The valence quark distribution is given by the even part of
Eq.(\ref{q_uni}),
\be
u(x) + d(x) - \bar u(x) - \bar d(x)
&=& N_c [ f_{\rm lev}(x) + f_{\rm lev}(-x)] 
\nonumber \\
&+& N_c \sum_{\rm neg.\, cont.} 
[ f_n (x) + f_n (-x) - f_n^{(0)}(x) - f_n^{(0)}(-x) ] .
\label{q_val}
\ee
Here the continuum contribution is finite and does not require an UV
cutoff.  This distribution is normalized by the baryon number sum rule,
\be
\int_0^1 dx\, [u(x) + d(x) - \bar u (x) - \bar d (x)] &=& N_c ,
\label{number_sum_rule}
\ee
which holds ``level by level'' in the chiral soliton model, as can be seen
by integrating Eq.(\ref{q_uni_level}) over $x$, upon which the
$\gamma^0\gamma^3$--term drops out.
\par
The total distribution of quarks plus antiquarks is given by the odd part
of Eq.(\ref{q_uni}). The continuum contribution to this function is
logarithmically divergent for each value of $x$ and requires
regularization. (Again, the divergence is proportional to $M^2$
\cite{DPPPW97}.) We regularize it by a Pauli--Villars subtraction,
analogous to the energy, Eq.(\ref{E_tot_reg}):
\be
\lefteqn{u(x) + d(x) + \bar u(x) + \bar d(x) 
\;\; = \;\; N_c [ f_{\rm lev}(x) - f_{\rm lev}(-x)] } && \nonumber \\
&+& N_c \sum_{\rm neg.\, cont.} 
[ f_n (x) - f_n (-x) - f_n^{(0)}(x) + f_n^{(0)}(-x) ] 
\nonumber \\
&-& N_c \frac{M^2}{M_{PV}^2}
\sum_{\rm neg.\, cont.} 
[ f_{PV, n} (x) - f_{PV, n} (-x) - f_{PV, n}^{(0)}(x) + 
f_{PV, n}^{(0)}(-x) ] .
\label{q_tot_reg}
\ee
Note that, again, this regularization differs from the one used in
refs.\cite{DPPPW96,DPPPW97} in that only the Dirac continuum is subtracted
here but not the level contribution. We emphasize that such regularization
preserves the general properties of parton distributions just as the one
used in refs.\cite{DPPPW96,DPPPW97}; in particular, it satisfies all
criteria for a ``good'' regularization established in ref.\cite{DPPPW97},
such as absence of anomalies, uniform logarithmic cutoff dependence, {\em
etc.}
\par
The total distribution of quarks plus antiquarks regularized according to
Eq.(\ref{q_tot_reg}) satisfies the momentum sum rule, if the classical pion
field is a stationary point of the regularized energy,
Eq.(\ref{E_tot_reg})\footnote{We note that for Eq.(\ref{msr}) to be
satisfied it is also necessary that $M_N$ in Eq.(\ref{q_uni_level}) be the
minimum value of the static energy, {\em i.e.}, the nucleon mass in the
leading order of the $1/N_c$--expansion quoted above.}:
\be
\int_0^1 dx\, x [u(x) + d(x) + \bar u (x) + \bar d (x)] \;\; 
= \;\; 1.
\label{msr}
\ee
A general proof of the momentum sum rule has been given in
ref.\cite{DPPPW96}, which can straightforwardly be carried over to the
Pauli--Villars regularized energy, Eq.(\ref{E_tot_reg}), and distribution
function, Eq.(\ref{q_tot_reg}).  Note that the momentum sum rule differs
from the baryon number sum rule in that it is a statement only about the
sum over all occupied levels but does not hold ``level by level''.
\par
{\em Results.} We compute the distribution functions Eqs.(\ref{q_val},
\ref{q_tot_reg}) for the self--consistent soliton profile using the
numerical method developed in ref.\cite{DPPPW97}.  For the numerical
calculations it is convenient to make use of the fact that with
Pauli--Villars regularization one may equivalently compute the distribution
function at negative $x$ by summing over non-occupied states, thus avoiding
the need of vacuum subtraction:
\be
f_{\rm lev}(-x) + \sum_{\rm neg.\, cont.} f_n (-x) 
&\rightarrow& - \sum_{\rm pos.\, cont.} f_n (x) ,
\label{occ_nonocc}
\ee
and similarly for the sum with $M \rightarrow M_{PV}$; see
ref.\cite{DPPPW97} for details.
\par
\begin{figure}
\vspace{-1cm}
\epsfxsize=16cm
\epsfysize=15cm
\epsffile{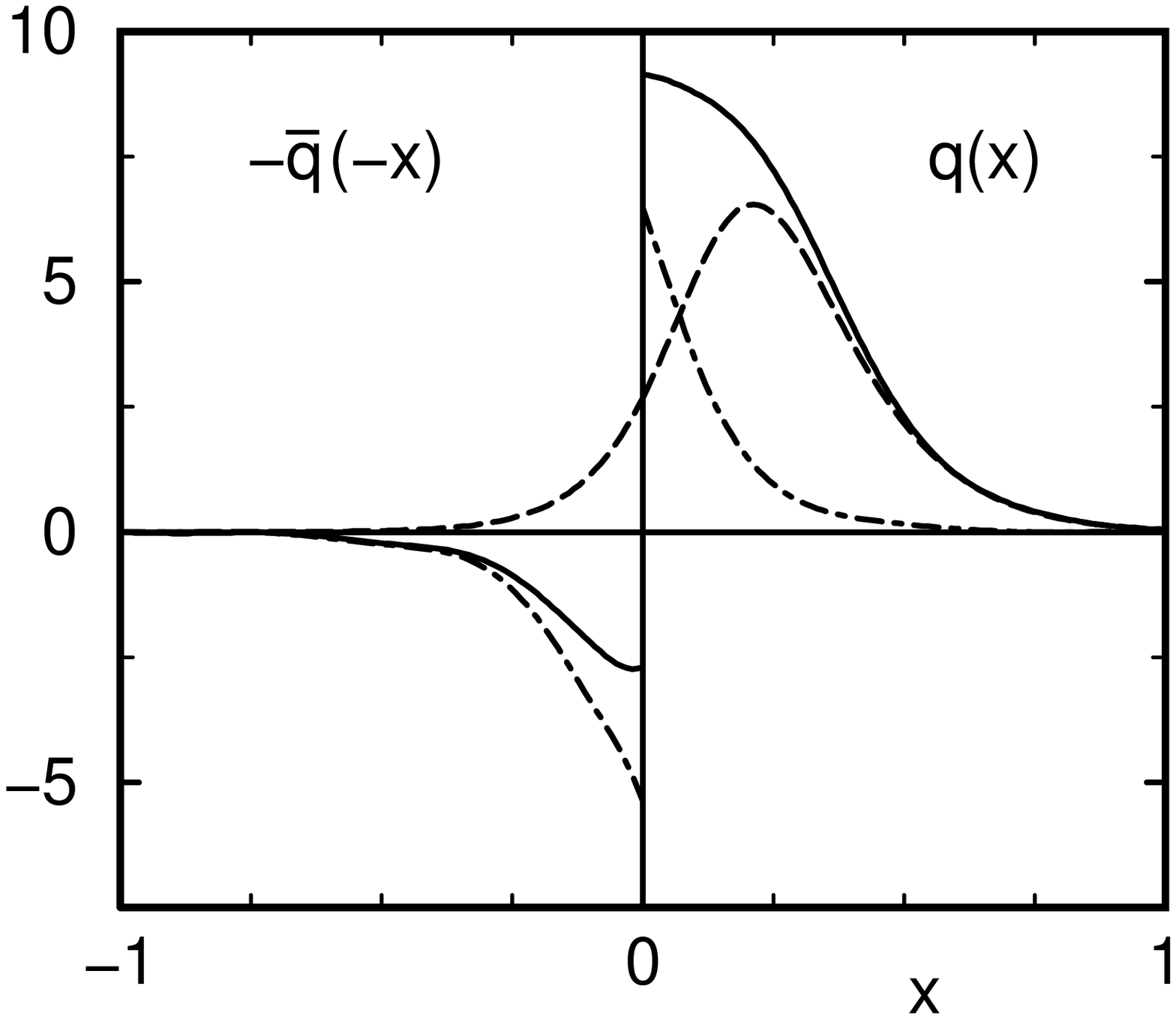}
\caption[]
{The isosinglet distribution function, Eq.(\ref{q_uni}), corresponding to
$u(x) + d(x)$ for $x > 0$ and $-\bar u(-x) - \bar d(-x)$ for $x < 0$, for
the self--consistent soliton with $M = 350\, {\rm MeV}$.  {\em Dashed
line:} contribution of the bound--state level, $f_{\rm lev} (x)$, giving a
negative contribution to the antiquark distribution. {\em Dot--dashed
line:} contribution of the Pauli--Villars regularized Dirac continuum.
{\em Solid line:} total result (bound--state level plus Dirac
continuum). The total antiquark distribution is positive.}
\label{fig_q_uni}
\end{figure}

The distribution functions receive contributions from the bound--state
level as well as from the polarized Dirac continuum of quarks.  To
understand the nature of these contributions it is instructive to consider
for a moment the ``universal'' function, Eq.(\ref{q_uni}), which describes
the quark distribution at positive and {\em minus} the antiquark
distribution at negative values of $x$.  Fig.\ref{fig_q_uni} shows the
contributions of the bound--state level and the Dirac continuum to this
function ($M = 350\,{\rm MeV}$), the latter being regularized by a
Pauli--Villars subtraction as in Eq.(\ref{q_tot_reg}).  The distribution
resulting from the bound--state level (dashed line) is positive for all
values of $x$; hence it makes a {\em negative} contribution to the
antiquark distribution. The correct sign of the antiquark distribution is
obtained by including the contribution of the Dirac continuum (dot--dashed
line), which is positive for $x > 0$ and negative for $x < 0$, in agreement
with the fact that the logarithmic divergence of the distribution function
is an odd function $x$.  We note that the so-called valence quark
approximation advocated in ref.\cite{WGR97}, when applied to parton
distribution functions, would violate the positivity of the antiquark
distributions.
\par
\begin{figure}
\vspace{-1cm}
\epsfxsize=16cm
\epsfysize=15cm
\epsffile{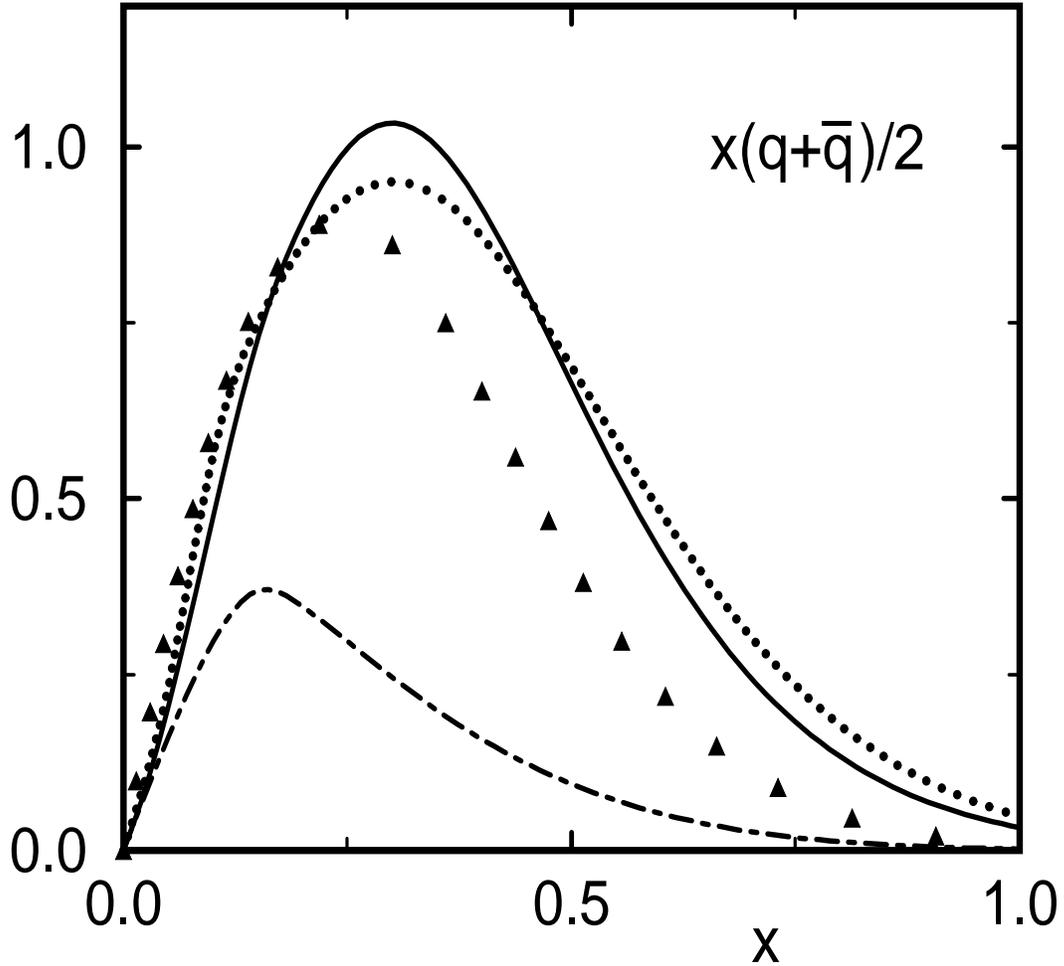}
\caption[]
{The isosinglet distribution of quarks plus antiquarks, 
$\frac{1}{2} x [u(x) + d(x) + \bar{u}(x) + \bar{d}(x)]$. Shown are the
total result ({\em solid line}) and the contribution of the Dirac continuum
({\em dot--dashed line}) for the self--consistent soliton with 
$M = 350\, {\rm MeV}$. Also shown is the total result for the soliton with
$M = 420\, {\rm MeV}$ ({\em dotted line}). {\em Triangles:} the
distribution calculated in ref.\cite{DPPPW97} with an analytic soliton
profile and regularized level contribution.}
\label{fig_q_tot}
\end{figure}

The total distribution of quarks plus antiquarks, 
$u(x) + d(x) + \bar u (x) + \bar d (x)$, is obtained by taking the odd part
in $x$ of the function shown in Fig.\ref{fig_q_uni}, or, equivalently,
performing the sum Eq.(\ref{q_tot_reg}). The resulting distribution is
shown in Fig.\ref{fig_q_tot} (solid line). It receives a sizable
contribution from the Dirac continuum (dot--dashed line). We emphasize that
the calculated distribution explicitly satisfies the momentum sum rule,
thanks to the self--consistency of the classical pion field.
\par
\begin{figure}
\vspace{-1cm}
\epsfxsize=16cm
\epsfysize=15cm
\epsffile{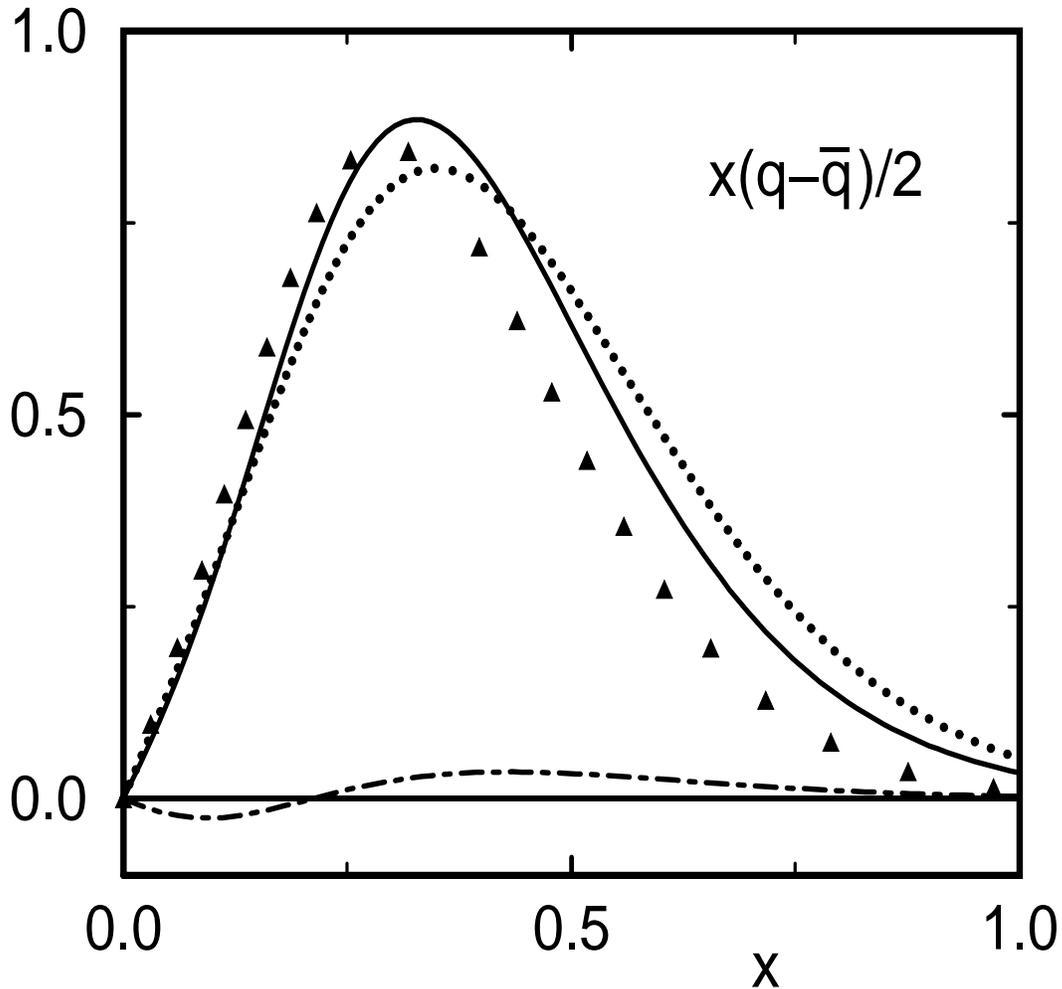}
\caption[]
{The isosinglet valence quark distribution, 
$\frac{1}{2} x [u(x) + d(x) - \bar{u}(x) - \bar{d}(x)]$.  Shown are the
total result ({\em solid line}), and the contribution of the unregularized
Dirac continuum ({\em dot--dashed line}) for the self--consistent soliton
with $M = 350\, {\rm MeV}$.  This distribution is obtained from the
symmetric part of the function in Fig.\ref{fig_q_uni} after removing the
Pauli--Villars cutoff, $M_{PV}/M \rightarrow \infty$. Also shown is the
total result for the soliton with $M = 420\, {\rm MeV}$ ({\em dotted
line}). {\em Triangles:} the distribution calculated in ref.\cite{DPPPW97}
with an analytic soliton profile.}
\label{fig_q_val}
\end{figure}

The valence quark distribution, $u(x) + d(x) - \bar u (x) - \bar d (x)$, is
shown in Fig.\ref{fig_q_val}. This distribution one obtains by taking even
part of the function shown in Fig.\ref{fig_q_uni} and removing the
Pauli--Villars regularization, $M_{PV}/M \rightarrow \infty$, or,
equivalently, by evaluating Eq.(\ref{q_val}). In this distribution the
logarithmic divergence of the Dirac continuum cancels; the finite continuum
contribution obtained in the limit $M_{PV}/M \rightarrow \infty$ is shown
in Fig.\ref{fig_q_val} (dot--dashed line); it is numerically small. The
calculated distribution satisfies the baryon number sum rule,
Eq.(\ref{number_sum_rule}).  Also shown in Figs.\ref{fig_q_tot} and
\ref{fig_q_val} (triangles) are the distributions calculated in
ref.\cite{DPPPW97} with Pauli--Villars subtraction of the level
contribution and a fixed soliton profile.
\par
\begin{figure}
\vspace{-1cm}
\epsfxsize=16cm
\epsfysize=15cm
\epsffile{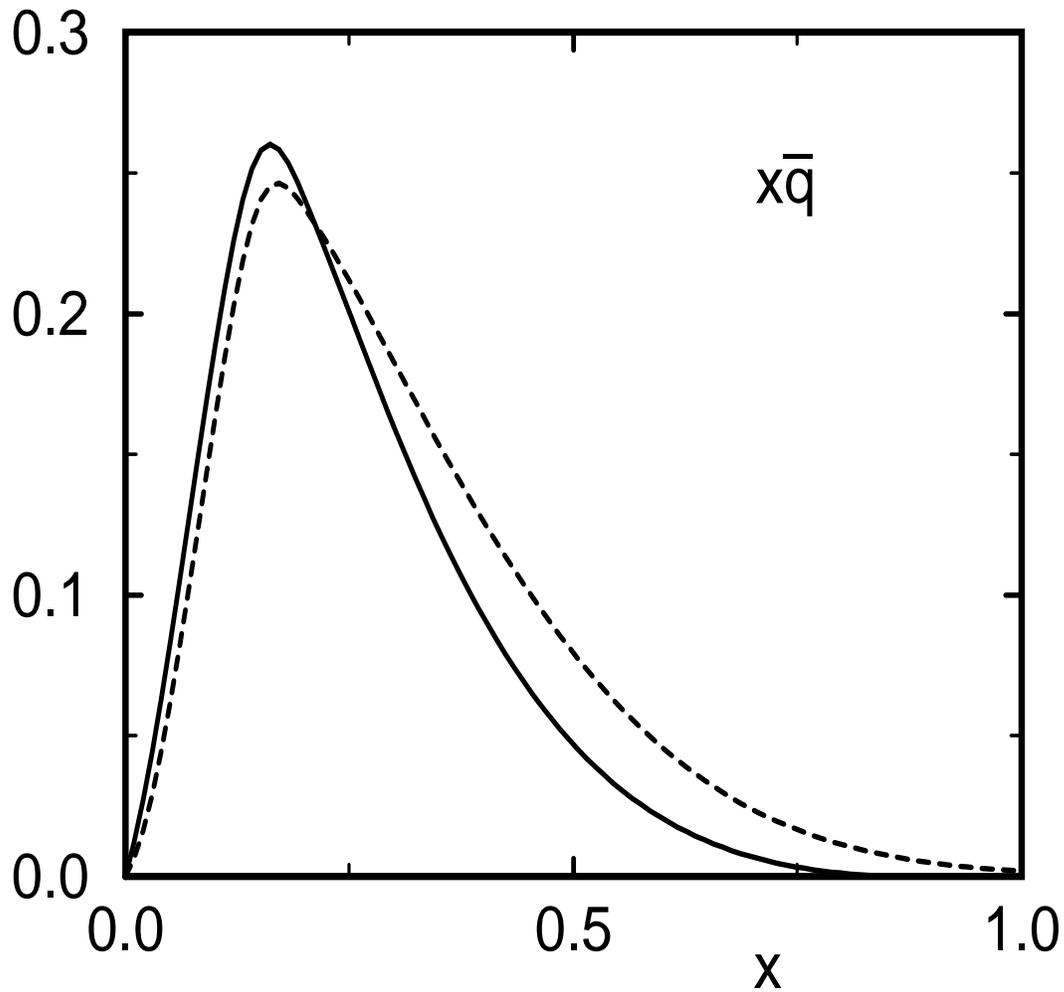}
\caption[]
{The isosinglet antiquark distribution, $x [\bar{u}(x) + \bar{d}(x)]$, for
the self-consistent soliton with $M = 350\, {\rm MeV}$, obtained as
difference between $u(x) + d(x) + \bar u(x) + \bar d(x)$,
Fig.\ref{fig_q_tot}, and $u(x) + d(x) - \bar u(x) - \bar d(x)$,
Fig.\ref{fig_q_val}, with different regularizations of the continuum
contribution to $u(x) + d(x) - \bar u(x) - \bar d(x)$. {\em Solid line:}
unregularized.  {\em Short--dashed line:} regularized with 
$M_{PV}^2/M^2 = 2.25$.}
\label{fig_q_ant}
\end{figure}

The valence quark distribution receives a non-zero contribution from the
Dirac continuum (Fig.\ref{fig_q_val}, dot--dashed line). The contribution
of the Dirac continuum to the baryon number, Eq.(\ref{number_sum_rule}),
however, is exactly zero. This is in fact necessary for the nucleon to have
baryon number unity, since the baryon number of the bound--state level is
already unity. A Pauli--Villars regularization of the Dirac continuum
contribution as in $u(x) + d(x) + \bar u (x) + \bar d (x)$,
Eq.(\ref{q_tot_reg}), would not violate the baryon number sum rule --- the
contributions of the true and the Pauli--Villars continuum to the baryon
number would be individually zero --- and thus constitute an admissible, if
``unnecessary'', finite regularization.  Since aside from preserving sum
rules and other general requirements the UV regularization of the effective
theory is fraught with some uncertainty, it is instructive to explore the
consequences of such finite regularization on the distribution
functions. For the valence distribution,
$u(x) + d(x) - \bar u (x) - \bar d (x)$, the influence of a regularization
of the Dirac continuum is negligible, since this distribution is dominated
by the bound--state level contribution, see Fig.\ref{fig_q_val}. The
antiquark distribution, however, which is defined as the difference of
Eq.(\ref{q_tot_reg}) and Eq.(\ref{q_val}), receives a large contribution
from the Dirac continuum and is potentially more sensitive to finite
regularizations of $u(x) + d(x) - \bar u (x) - \bar d (x)$.  In
Fig.\ref{fig_q_ant} we show the antiquark distribution obtained as the
difference between the same distribution of quarks plus antiquarks
(Fig.\ref{fig_q_tot}, $M = 350\, {\rm MeV}$) and a valence distributions
computed with {\em i)} unregularized continuum contribution (solid line)
and {\em ii)} continuum Pauli--Villars regularized with 
$M_{PV}^2/M^2 = 2.52$ as in the total distribution (short--dashed
line). One sees that the effect of the finite regularization on the
antiquark distribution is not dramatic.  In particular, the antiquark
distribution is positive in both cases.  Of course, it is guaranteed to be
positive in the limit $M_{PV}/M \rightarrow \infty$, because of the
logarithmic divergence of $u(x) + d(x) + \bar u (x) + \bar d (x)$. The fact
that it is positive also for finite $M_{PV} / M$ shows that the finite,
regularization--dependent terms do not upset this asymptotic statement for
the particular regularization used here, as it should be.
\par
Figs.\ref{fig_q_tot} and \ref{fig_q_val} also show the distributions
obtained for the soliton with $M = 420\,{\rm MeV}$ (dotted lines).  We see
that the distributions calculated with the self--consistent soliton profile
are rather insensitive to the value of the constituent quark mass.  This
parallels the behavior of most hadronic observables in this model
\cite{Review}.
\par
The numerical method employed here \cite{DPPPW97} does not allow for an
accurate determination of the limiting behavior of the parton distributions
for $x \rightarrow 0$. The limiting behavior of the Dirac continuum
contribution in Fig.\ref{fig_q_uni} has been obtained by extrapolation, see
ref.\cite{DPPPW97} for details; for the distributions multiplied by $x$,
Figs.\ref{fig_q_tot}, \ref{fig_q_val} and \ref{fig_q_ant}, this limitation
of the present method is irrelevant. We note also that the ``small--x''
behavior of the parton distributions in the effective chiral theory, {\em
i.e.}, the behavior in the parametrically small region 
$|x| \le (M /\Lambda )^2 / N_c$, depends on the details of the UV
regularization, as has recently been emphasized in connection with
off--forward parton distributions \cite{PPPBGW97}.
\par
The parton distributions studied here pertain to the large--$N_c$ limit,
where the nucleon is heavy, $M_N \propto N_c$. The calculated distributions
therefore do not go to zero at $x = 1$, rather, they are exponentially
small at large $x$, as discussed in ref.\cite{DPPPW96}.
\par
{\em Summary and Conclusions.} We have shown, by a specific example, that
the general requirements on the UV regularization in the calculation of
parton distributions in the effective chiral theory can be reconciled with
the requirements of a stable soliton solution, which depends on 
$O(M^2 / \Lambda^2)$--terms in the energy functional. The calculated quark
and antiquark distributions explicitly satisfy the baryon number and
momentum sum rules.
\par
The distributions obtained with the self--consistent soliton profile
largely support the results of the calculations of
refs.\cite{DPPPW96,DPPPW97} using a fixed soliton profile. Except for the
momentum sum rule the differences are not significant, given the limited
theoretical accuracy of this approach.  We note that the regularization
used in the present work is only an example of a class of possible
regularizations leading to a stable soliton; for instance, one may perform
multiple Pauli--Villars subtraction of the Dirac continuum with different
masses, {\em etc}. However, we see at present no need to explore such
possibilities.
\par
The basic expressions for the parton distributions in
refs.\cite{DPPPW96,DPPPW97} have been derived in the leading order of
$M/\Lambda$, in which the ``constituent'' quarks of the effective theory
can be regarded as pointlike. This is why the momentum sum rule is
saturated by the quark and antiquark distributions computed here.  If one
knew the precise form of the effective theory at order $M^2 /\Lambda^2$,
one could ``resolve'' the structure of the constituent quark and recover
the distribution of quarks and gluons inside the constituent quark. Such a
picture is in fact suggested by the instanton vacuum, where the parameter
$M^2 /\Lambda^2$ is related to the small packing fraction of the instanton
medium \cite{DPW96,BPW97}; its concrete realization is the subject of
current investigations.  It is not unlikely that the ``resolution'' of the
constituent quark structure could be represented in the form of a
convolution in Bjorken--$x$ of the constituent quark distributions with
parton distributions inside the constituent quark.  This general idea is in
fact rather old \cite{ACMP74}; it has recently been taken up again
\cite{Scopetta97}. Since the ``constituent'' quark and antiquark
distributions computed here are correctly normalized to the total nucleon
momentum, they can serve as a starting point for investigations assuming a
structure inside the constituent quark.
\par
{\em Acknowledgements.} The authors are indebted to D.I.\ Diakonov, V.Yu.\
Petrov, P.V.\ Pobylitsa and M.V.\ Polyakov for numerous helpful
suggestions.
%

%
%
\newpage
\newpage
\newpage
\newpage
\end{document}